\documentstyle[12pt,epsf]{article} 
\newcommand{\subs}[1]{\subsection{#1}\setcounter{equation}{0}}
 
\newcommand{\Tr}{{\rm Tr}\;}

\newcommand{\half}{\small{1\over 2}} 
\newcommand{\vac}{|0\rangle} 

\font\zfont = cmss10 scaled\magstep1 
\def\ZZ{\hbox{\zfont Z\kern-.4emZ}}
\def\gom#1{\gamma_{\Omega{#1}}} 
\def\gomr#1#2{\gamma_{R_{#1}\Omega{#2}}}
\def\gomt#1{\gamma^T_{\Omega{#1}}}
\def\gomrt#1#2{\gamma^T_{R_{#1}\Omega{#2}}} 

\newcommand{\npb}[3]{Nucl. Phys. {\bf B#1} (#2) #3}

\newcommand{\prl}[3]{Phys. Rev. Lett. {\bf #1} (#2) #3}
\newcommand{\mpl}[3]{Mod. Phys. Lett. {\bf A#1} (#2) #3}

\newcount\yearltd\yearltd=\year\advance\yearltd by -1900
\newcommand{\eqn}[2]{\begin{equation}\label{eq:#1}#2\end{equation}} 
\newcount\figurecount	\figurecount=0
\newskip\captionskip	\captionskip=15pt plus 5pt minus 3pt
\newdimen\captionwidth	\captionwidth=\hsize
\font\captionfont = cmr10 
\newcommand\figcaption[1]{
	\vbox{\hsize=\the\captionwidth%
		\vskip\the\captionskip \noexpand{\captionfont
		Figure~\the\figurecount:\enspace #1}
	}}
\newcommand{\efig}[3]{
	\begin{figure}
	\global\advance\figurecount by 1
	\begin{center}
	\vbox{\leavevmode\epsfysize=#1 \epsfbox{#2} \figcaption{#3}}
	\end{center}
	\end{figure}
}

\def\boxit#1{\vbox{\hrule\hbox{\vrule\kern3pt
\vbox{\kern3pt#1\kern3pt}\kern3pt\vrule}\hrule}}
\newdimen\str
\def\fboxit#1#2{\vbox{\hrule height #1 \hbox{\vrule width #1
\kern3pt \vbox{\kern3pt#2\kern3pt}\kern3pt \vrule width #1 }
\hrule height #1 }}
\def\yboxit#1#2{\vbox{\hrule height #1 \hbox{\vrule width #1
\vbox{#2}\vrule width #1 }\hrule height #1 }}
\def\fillbox#1{\hbox to #1{\vbox to #1{\vfil}\hfil}}
\def\dotbox#1{\hbox to #1{\vbox to 8pt{\vfil}\hfil $\cdots$ \hfil}}
\def\ybox{\yboxit{0.4pt}{\fillbox{8pt}}\hskip-0.4pt}

\def\yboxs{\yboxit{0.4pt}{\fillbox{5pt}}\hskip-0.4pt}

\def\tableaux#1{\vcenter{\offinterlineskip \halign{&\tabskip 0pt##\cr
#1}}\ }
\def\cropen#1{\crcr\noalign{\vskip #1}}
\def\cry{\cropen{-0.4pt}}
\def\fnote#1#2{\begingroup\def\thefootnote{#1}\footnote{#2}
\addtocounter{footnote}{-1}\endgroup}

\begin{document}\bigskip

\begin{titlepage}

%\begin{flushleft}
%\draftdate\\
%\end{flushleft}

\begin{flushright}
RU-96-28\\
hep-th/9605049
\end{flushright}

\vskip 1.5in
\begin{center}
{\Large \bf A D=4 N=1 Orbifold of Type I Strings}

\vskip 1cm
{
Micha Berkooz\fnote{*}{e-mail: {\tt berkooz@physics.rutgers.edu}}\
and Robert G. Leigh\fnote{\dag}
{e-mail: {\tt leigh@physics.rutgers.edu}}\\
Department of Physics\\
Rutgers University\\
Piscataway, NJ\ \ 08855-0849
}

\end{center}

\begin{abstract} 
\baselineskip=16pt 
We consider the propagation of Type I open superstrings 
on orbifolds with four non-compact dimensions and $N=1$
supersymmetry. In this paper,
we concentrate on a non-trivial $\ZZ_2\times \ZZ_2$ example.
We show that consistency conditions, arising from tadpole
cancellation and algebraic sources, require the existence of three
sets of Dirichlet 5-branes. We discuss fully the 
enhancements of the spectrum when these 5-branes intersect.
An amusing attribute of these models is the importance of
the tree-level (in Type I language) superpotential to
the consistent relationship between Higgsing and the motions of
5-branes.
\end{abstract} 

\vfill 6 May 1996

\end{titlepage}

\newpage 
\baselineskip=18pt

\subs{Introduction}

The solution of $N=1$ field theories in four dimensions relies 
heavily on understanding different limits of moduli space.
One certainly expects that such a tool will play a significant 
role in string theories with $N=1$ supersymmetry as well.
Strong-weak coupling duality of string theory allows us to control
the strong coupling regime of some string theories in terms of
other weakly coupled ones. One such example is Type I/heterotic
duality;  one of the first steps towards solving the low 
energy dynamics of the $SO(32)$ heterotic string theory is to understand 
Type I compactifications with $N=1$ supersymmetry. 
In this paper we will discuss
a simple but non-trivial compactification of Type I strings to four
dimensions.

There are other motivations for studying four dimensional
compactifications of the Type I theory. One aspect is a potential
importance of this technology for phenomenology: for example, one 
finds matter fields
which transform as bilinears of the fundamental representation in
vast quantities (open strings have two ends). We may also hope to
find, within a general Type II orientifold framework, constructions with
chiral fermions, although the model discussed here does not have
this property. Lastly, the couplings
in Type I theory (or open string sectors in general) may behave 
quite differently than expected from
perturbative heterotic string theory, alleviating 
some unpleasant general features of heterotic string theory model 
building\cite{strcy}.

Another motivation is the study of solitons in string theory. 
Free or solveable conformal field theories 
that admit solitons or other non-perturbative objects 
provide the most reliable avenue for discussing the properties of 
these objects.

Type I superstrings compactified to {\em six} dimensions on a 
K3 orbifold have
been considered in several  recent papers. In Ref. \cite{gimpol}, the
worldsheet consistency conditions were studied. It was found that for
consistent open string propagation, there must be 32 parallel 
5-branes,\cite{dbrane}
as well as the 32 9-branes found in 10-dimensional theory\cite{polcai}.
Furthermore, spacetime anomaly constraints were studied in Ref.
\cite{blpss}, and it was found that parts of the gauge groups found in
Ref. \cite{gimpol} were in fact broken through a modification of the
Green-Schwarz mechanism. Other six-dimensional orientifolds have
been discussed in  Refs. \cite{otherkthree}.

In this paper, we begin a study of orientifolds with four non-compact
dimensions. Here we discuss Type I superstrings compactified on
$T^6/\ZZ_2\times\ZZ_2$, where the two $\ZZ_2$'s act as follows on the
compact coordinates: 
\begin{eqnarray} 
\label{eq:orbifold} 
R_1&:&
X^{6,7,8,9} \rightarrow -X^{6,7,8,9} \nonumber\\ 
& & X^{4,5} \rightarrow +X^{4,5} \\ 
R_2&:& 
X^{4,5,8,9} \rightarrow -X^{4,5,8,9} \nonumber\\
& & X^{6,7} \rightarrow +X^{6,7} \nonumber 
\end{eqnarray} 
The orbifold group also includes the projection $R_3=R_1R_2$ which acts
on the $4,5,6,7$ coordinates. Since we consider Type I strings, we will
consider the worldsheet orientation reversal $\Omega$ as well. Thus the
full orientifold group contains the elements 
$\{ 1,R_i,\Omega,\Omega R_i\}, i=1,2,3$.

This is a seemingly complicated orbifold to begin with, but it is
perhaps the simplest non-trivial extension of Ref. \cite{gimpol} to four
dimensions, and as we will see has some very interesting new features.
We construct here a consistent Type I compactification and study
the tree-level superpotential and Higgsing phenomena in some detail.
Many details of this and other models will be left for further
publications. In particular, there are many interesting non-perturbative
aspects to be explored.

In the next section, we discuss some general aspects of the
spectrum of these theories. In the following section,
we will give some details of the worldsheet
consistency conditions following from tadpole cancellation. 
We will find that three orthogonal sets of 32 5-branes will be
necessary, in addition to 32 9-branes. This configuration preserves
$N=1$ supersymmetry in four
dimensions and leads to a rich variety of phenomena associated with
5-brane morphology. The remaining supersymmetry may be demonstrated as
follows. First, on the worldsheet, we take the action of $R_1$ and 
$R_2$ to be 
\begin{eqnarray}
\label{eq:fermorb} 
R_1&=&exp(i\pi (J_{67}+J_{89}))\\ 
R_2&=&exp(i\pi (J_{45}+J_{89}))\nonumber
\end{eqnarray} 
when acting on any worldsheet field. Thus the worldsheet
supercharge left invariant by these operations is given by
\begin{equation} 
\label{eq:superchg}
e^{-\phi/2}e^{i(H_0+H_1+H_2+H_3-H_4)/2} 
\end{equation}
One could worry that this does not look invariant under $R_3$ if defined
similarly to (\ref{eq:fermorb}); however, the identification
$R_3=R_1R_2$ leads to: 
\begin{equation}
\label{eq:fermorbthr} 
R_3=e^{2\pi iJ_{89}} exp(i\pi (J_{45}+J_{67})) 
\end{equation} 
The non-trivial $J_{89}$ factor makes the supercharge invariant. From
the spacetime point of view, we see the same result: the $\Omega$
projection leaves $Q+\tilde Q$ invariant. The $R_i$ projection leaves
$Q+R_i\tilde Q$; only those components with $R_i$-eigenvalues equal to
$+1$ correspond to unbroken supersymmetries. Since $R_3=R_1R_2$, there
are only two such independent conditions and thus the supersymmetry is
reduced by a factor of $1/8$ from what it would have been in the Type
IIB case (which would give $N=8$ in $D=4$).

In Section 3, we discuss consistency conditions arising
from the cancellation of unphysical Ramond tadpoles.
In Section 4, we discuss further
constraints and derive the spectrum of the theory in
different configurations of $5$-branes. Finally, in the
last sections, we derive the tree-level superpotential and
demeonstrate its importance to T-duality and the
correspondence between the motion of $5$-branes and Higgsing.
 
\subs{The Orientifold Group and the Spectrum}

We use throughout the notation of Ref. \cite{gimpol}.
The orientifold group $G=G_1+\Omega G_2$
acts on open string states as: 
\begin{eqnarray} 
\label{eq:orientifold} 
g: |\psi, ij\rangle &\rightarrow \left( \gamma_g\right)_{ii'} 
| g\cdot\psi, i'j'\rangle \left( \gamma_g^{-1}\right)_{j'j}\\ 
\Omega h: |\psi, ij\rangle &\rightarrow 
\left( \gamma_{\Omega h}\right)_{ii'} 
| \Omega h\cdot\psi, j'i'\rangle 
\left( \gamma_{\Omega h}^{-1}\right)_{j'j}\nonumber
\end{eqnarray} 
for $g\in G_1$ and $h\in G_2$. A great deal of effort will go into
obtaining a consistent representation for the various matrices
$\gamma$. These must form a (projective) representation of the
orientifold group and must pass several additional algebraic 
tests. First, operator products relate sectors to one another,
and (\ref{eq:orientifold}) must be consistent with this. Second,
the calculation of unphysical tadpoles will place further 
constraints. We will find that these constraints taken together
essentially determine the $\gamma$'s completely.
We begin with some remarks on the structure of the
spectrum. 

\subsubsection{Closed String Spectrum}

The untwisted states are formed out of: 
\begin{equation}
	\begin{array}{rlccccc} 
		&                       & R_1 & R_2 & R_3 && SO(2)_{ST}
	\nonumber\\ 
	NS:& \psi^{\mu}_{-1/2}\vac &  +  &  +  &  +  && \pm {\bf 1}
	\nonumber\\
	   & \psi^{4,5}_{-1/2}\vac &  +  &  -  &  -  && 2\times {\bf 0}
	\nonumber\\
	   & \psi^{6,7}_{-1/2}\vac &  -  &  +  &  -  && 2\times {\bf 0}
	\nonumber\\
	   & \psi^{8,9}_{-1/2}\vac &  -  &  -  &  +  && 2\times {\bf 0}
	\\
	 R:& |s_1s_2s_3s_4\rangle  &  +  &  +  &  +  && \pm {\bf 1/2}
	\nonumber\\
           &                       &  +  &  -  &  -  && \pm {\bf 1/2}
	\nonumber\\
           &                       &  -  &  +  &  -  && \pm {\bf 1/2}
	\nonumber\\
           &                       &  -  &  -  &  +  && \pm {\bf 1/2}
	\nonumber
	\end{array}
\end{equation}
We have listed the transformation of each state under the $R_i$,
as well as its representation under the spacetime Lorentz
group.
Because of $\Omega$, we must symmetrize in the NS-NS sector, and
antisymmetrize in the R-R sector. We thus get the spectrum (here
$m_1=4,5, m_2=6,7$, etc.):
\begin{equation}
	\begin{array}{rlcc}
	   &                       && SO(2)_{ST}
	\nonumber\\ 
	NSNS:& \psi^{\left(\mu\right.}_{-1/2}\vac_L \otimes
\tilde\psi^{\left.\nu\right)}_{-1/2}\vac_R && \pm {\bf 2} 
\oplus {\bf 0}
	\nonumber\\
	   & \psi^{\left(m_i\right.}_{-1/2}\vac_L \otimes
\tilde\psi^{\left. n_i\right)}_{-1/2}\vac_R && 9\times {\bf 0}
	\nonumber\\
	 RR:& \Psi_{\left[ \alpha \pm \pm'\right.}\otimes
\tilde\Psi_{\left. \beta\right] \pm \pm'} && 4\times {\bf 0}
	\nonumber\\
	RNS:& \psi^{\mu}_{-1/2}\vac_L \otimes
\tilde\Psi_{\alpha ++} && \pm {\bf 3/2} \oplus \pm {\bf 1/2}
	\\
	    & \psi^{m_1}_{-1/2}\vac_L \otimes
\tilde\Psi_{\alpha +-} && 2\times\pm {\bf 1/2}
	\nonumber\\
	    & \psi^{m_2}_{-1/2}\vac_L \otimes
\tilde\Psi_{\alpha -+} && 2\times\pm {\bf 1/2}
	\nonumber\\
	    & \psi^{m_3}_{-1/2}\vac_L \otimes
\tilde\Psi_{\alpha --} && 2\times\pm {\bf 1/2}
	\nonumber
	\end{array}
\end{equation}
where $\Psi_{\alpha\pm\pm'}$ labels the Ramond ground state
with helicity $\alpha$ and $R_1 (R_2)$ numbers $\pm (\pm')$. The
$\Omega$ projection symmetrizes the R-NS states given here with 
those of the NS-R sector. Thus the untwisted closed string sector
consists of the gravity multiplet, the dilaton chiral multiplet and six
chiral multiplets associated with the torus.

Next, we have states from the twisted sectors. Consider first the sector
twisted by $R_1$. The massless states are formed from :
\begin{equation}
	\begin{array}{rlcccc}
	   &                          & R_1 &    R_2   && SO(2)_{ST}
	\nonumber\\ 
	NS:& |s_3s_4\rangle, s_3=-s_4 &  +  & i(2s_4) && {\bf 0}
	\\
	 R:& |s_1s_2\rangle, s_1=-s_2 &  +  & i(2s_2) && \pm {\bf 1/2}
	\nonumber
	\end{array}
\end{equation}
We thus get the states:
\begin{equation}
	\begin{array}{rlcc}
	   &                       && SO(2)_{ST}
	\nonumber\\ 
	NSNS:& {1\over\sqrt{2}}\left( |+-\rangle\otimes |-+\rangle
	+|-+\rangle\otimes |+-\rangle\right) && {\bf 0}
	\nonumber\\
	  RR:& {1\over\sqrt{2}}\left( |+-\rangle\otimes |-+\rangle
	-|-+\rangle\otimes |+-\rangle\right) && {\bf 0}
	\nonumber\\
	RNS:& |+-\rangle\otimes |-+\rangle && {\bf 1/2}
	\\
	    & |-+\rangle\otimes |+-\rangle && -{\bf 1/2}
	\nonumber
	\end{array}
\end{equation}
where again, R-NS should be symmetrized with NS-R. We see then that we
get one chiral multiplet per twisted sector; this is the blowing up mode
for the corresponding fixed point. Each of the $R_i$ have 16 fixed
`points' (actually complex lines), so there are a total of 48 such chiral
multiplets.

\subsubsection{Open string states}

Consider first the 99-sector. In the NS sector, there are states of the
form $\psi^\mu|0,ab\rangle \lambda^0_{ab}$ and 
$\psi^{m_i}|0,ab\rangle \lambda^{(i)}_{ab}$. 
The Chan-Paton matrices satisfy:
\begin{eqnarray}\label{eq:open_nn}
\lambda^{(0)}=+\gamma_{R_1,9}\lambda^{(0)}\gamma_{R_1,9}^{-1}\;\; ;
\lambda^{(0)}=+\gamma_{R_2,9}\lambda^{(0)}\gamma_{R_2,9}^{-1}\;\; ;
\lambda^{(0)}=-\gamma_{\Omega,9}\lambda^{(0)T}\gamma_{\Omega,9}^{-1}
\nonumber\\
\lambda^{(1)}=+\gamma_{R_1,9}\lambda^{(1)}\gamma_{R_1,9}^{-1}\;\; ;
\lambda^{(1)}=-\gamma_{R_2,9}\lambda^{(1)}\gamma_{R_2,9}^{-1}\;\; ;
\lambda^{(1)}=-\gamma_{\Omega,9}\lambda^{(1)T}\gamma_{\Omega,9}^{-1} \\
\lambda^{(2)}=-\gamma_{R_1,9}\lambda^{(2)}\gamma_{R_1,9}^{-1}\;\; ;
\lambda^{(2)}=+\gamma_{R_2,9}\lambda^{(2)}\gamma_{R_2,9}^{-1}\;\; ;
\lambda^{(2)}=-\gamma_{\Omega,9}\lambda^{(2)T}\gamma_{\Omega,9}^{-1}
\nonumber\\
\lambda^{(3)}=-\gamma_{R_1,9}\lambda^{(3)}\gamma_{R_1,9}^{-1}\;\; ;
\lambda^{(3)}=-\gamma_{R_2,9}\lambda^{(3)}\gamma_{R_2,9}^{-1}\;\; ;
\lambda^{(3)}=-\gamma_{\Omega,9}\lambda^{(3)T}\gamma_{\Omega,9}^{-1}
\nonumber \end{eqnarray}

As indicated above, there will be three sets of 5-branes, which we will 
refer to as $5_i$-branes; the $5_i$-brane fills 4-dimensional spacetime
plus the $i^{th}$ $T^2$ spanned by $X^{m_i}$.
For 5-branes at fixed points, the $5_i5_i$-sector will satisfy
constraints similar to those of the 99-sector, except for a sign change 
in the $\Omega$-transformation for
the $\psi^{m_j}$ states ($j\neq i$).

Now consider moving the $5_i$-brane away from the $R_i$ fixed point. A
general configuration is shown (for $i=1$) in Fig. 1. Each 5-brane has
three images generically, and thus there can be at most eight together.
In this case, $R_i$ relates states to those of an image. $\Omega$ acts
as $-1$ (as well as transposing) on the six orthogonal dimensions and as
$+1$ on the other four. 
\efig{2in}{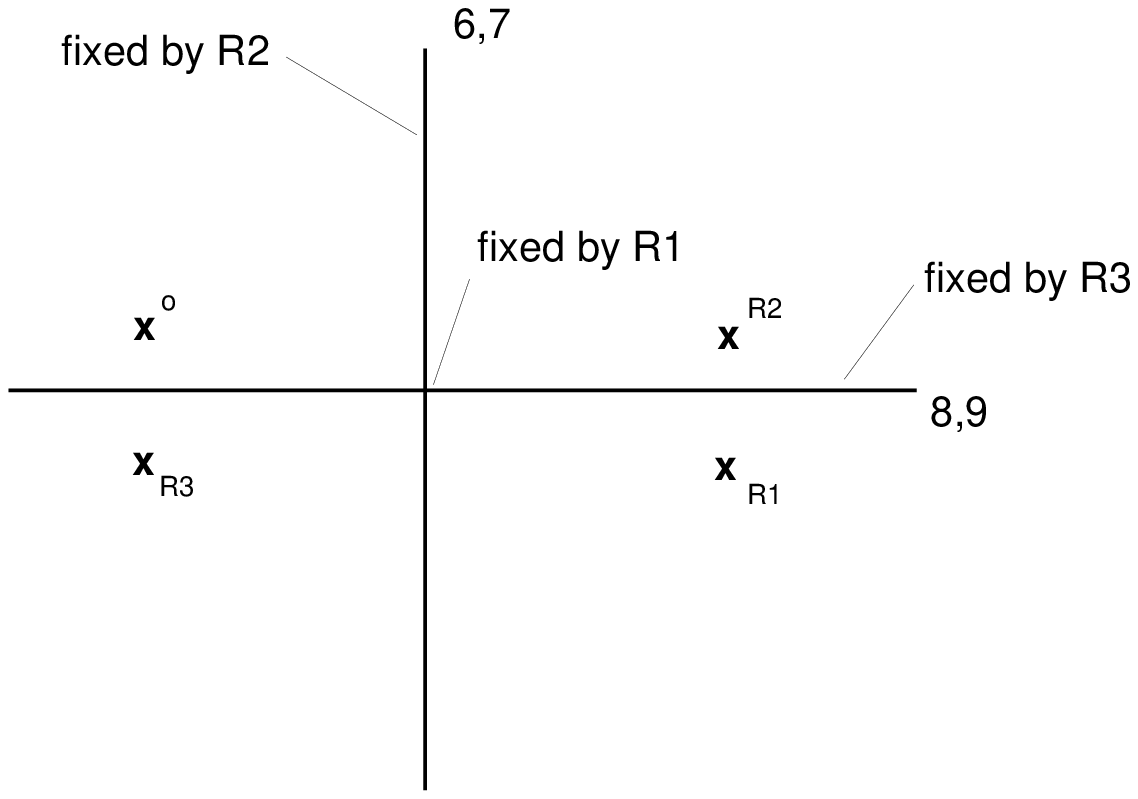}{Generic configuration of $5_1$-branes.}

There are also (complex) axes on which the $5_i$-brane is at a fixed
point of $R_j$ ($j\neq i$). As is clear from the figure, this 
corresponds to two of the images approaching each other across
the axis and there will then be extra massless states arising 
from strings stretching from one brane to an image.

Now let us discuss $5_i9$-states and $5_i5_j$-states more carefully. The
$5_i9$ states are much as in Ref. \cite{gimpol}. Consider a
$5_19$-state; since $X^{6,7,8,9}$ satisfy Neumann boundary conditions on
one end and Dirichlet on the other, these $X$'s will have $1/2$-integer
modings. The corresponding fermions have integer modings and thus the
vacuum forms a representation of the corresponding zero-mode Clifford
algebra. The NS state 
is\footnote{Note that the form of the supercharge (\ref{eq:superchg})
implies that there will be some sign changes in the GSO projection
in the $5_39$ sector.}
\begin{equation} 
5_19:\;\; |s_3s_4,ij\rangle\lambda_{ij},\; s_3=-s_4, 
\end{equation} 
constituting two real bosons (per element of $\lambda$). Away from the
fixed points, there are no constraints, apart from the GSO. The
$\Omega$-projection relates $5_i9$ to $95_i$ and the $R_i$-projection
relates one state to its image. The $R_j$-projections ($j\neq i$) also
map to an image. Now, if the $5_1$-brane is at a fixed point of $R_1$,
then $\lambda$ is restricted by 
$\lambda=\gamma_{R1,I}\lambda \gamma_{R1,I}^{-1}$. 
$R_j$ ($j\neq 1$) flip the sign of $X^{4,5}$; since
the lowest lying states have no dependence on $X^{4,5}$ (e.g., momentum
in these directions would cost energy), $\lambda$ is further restricted:
\begin{eqnarray}
\label{eq:fivninproj} 
\lambda=(2s_4i)\gamma_{R2}\lambda\gamma_{R2}^{-1}\\ 
\lambda=(2s_4i)\gamma_{R3}\lambda\gamma_{R3}^{-1} 
\end{eqnarray} 
where the phases are
deduced from eqs. (\ref{eq:fermorb},\ref{eq:fermorbthr}).

$5_i5_j$-states have some similarities to the $59$ states. Consider for
definiteness the $5_15_3$ state. A generically positioned $5_3$-brane 
appears as in Figs. 2. 
\efig{2in}{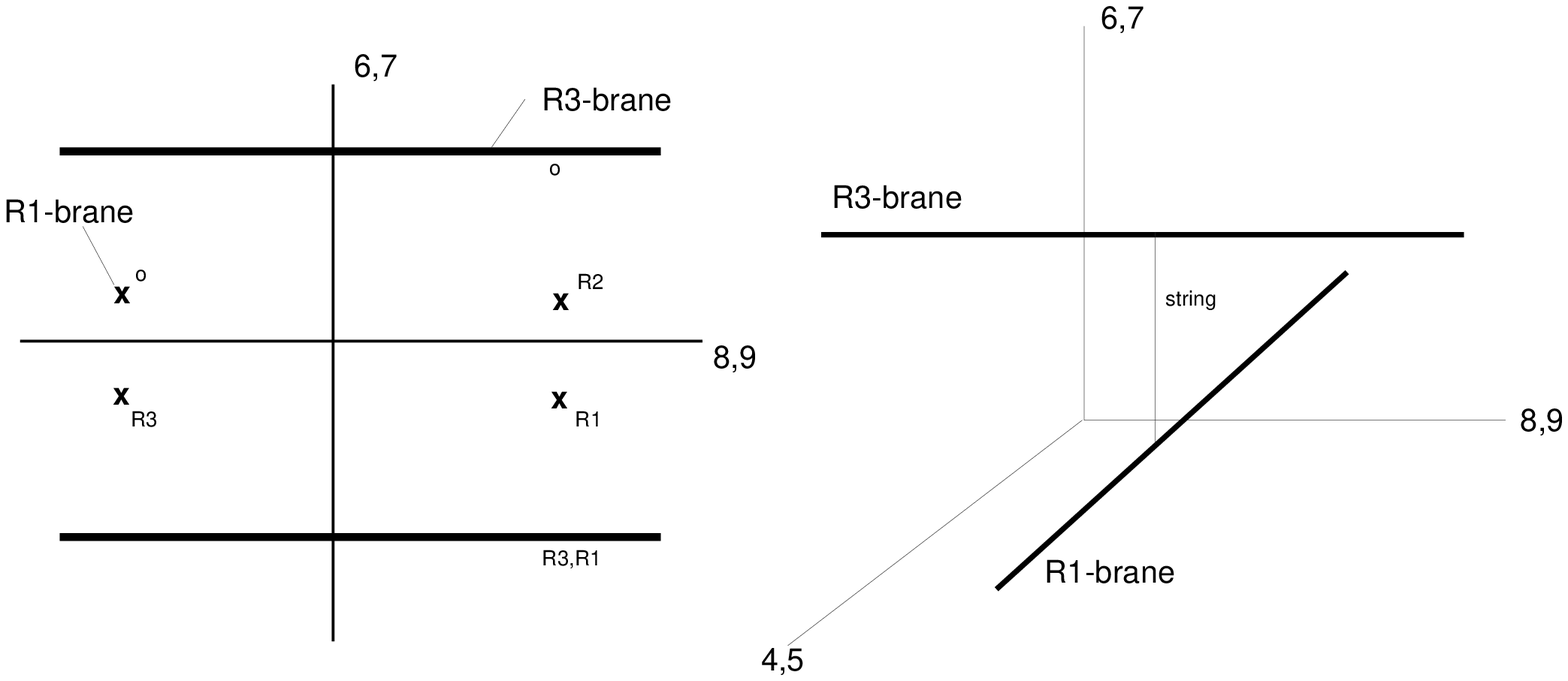}{Two views of a generic
configuration of $5_1$- and $5_3$-branes.} 
Clearly there can be massless
$5_15_3$-states if the $5_3$-brane in Figures 2 overlaps the
$5_1$-brane, i.e., they have the same $X^{6,7}$, as in Figure 3.
\efig{1.5in}{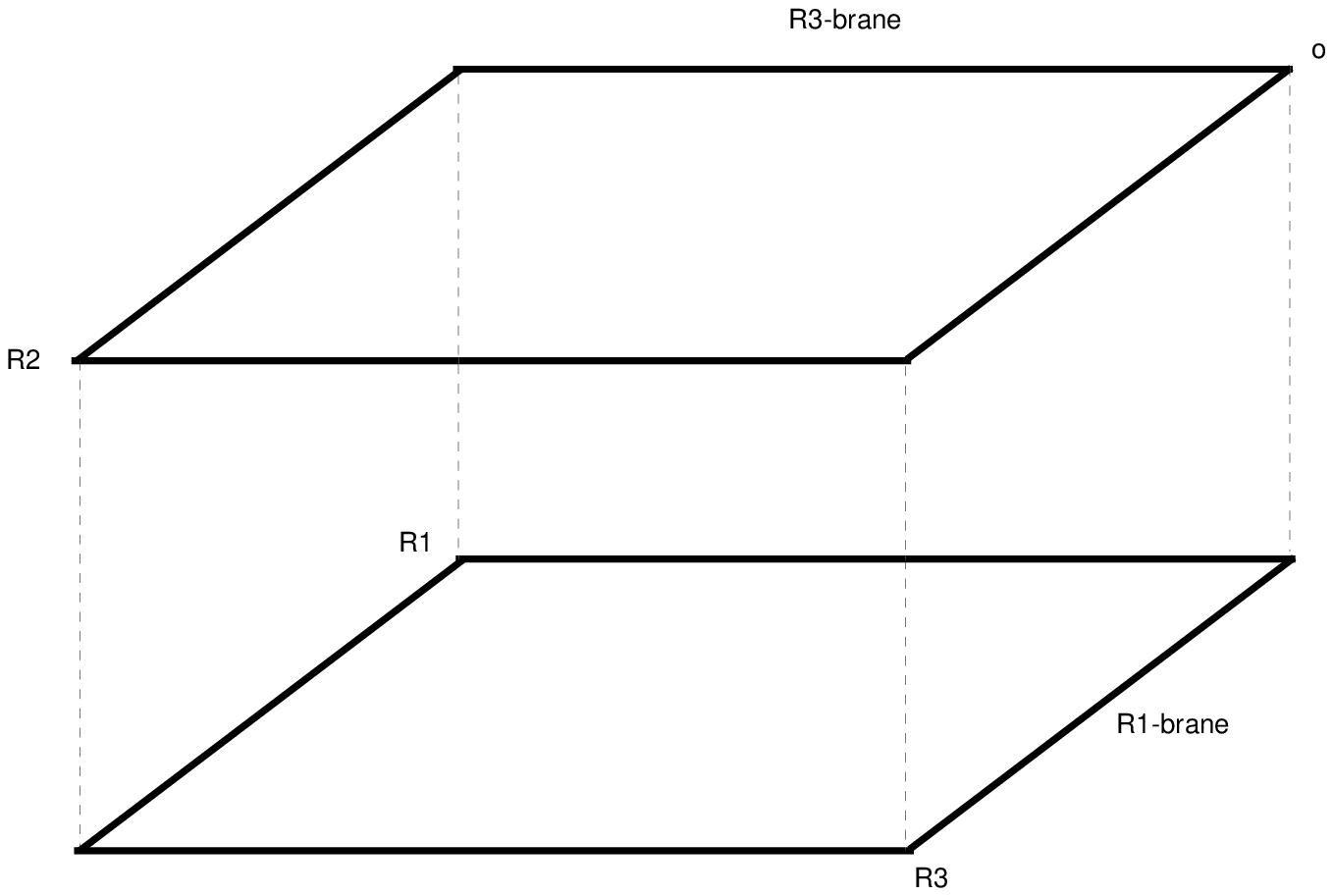}{Overlapping configuration of $5_1$- and
$5_3$-branes and their images.} 
In this case, the states will be of the
form $|s_2s_4,ij\rangle\lambda_{ij}$. There will be no restriction,
apart from GSO ($s_2=-s_4$) when the branes are at generic points, as
$R_i$ will map states to images. 
When symmetries are enhanced, for
example, if the $5_1$-brane intersects the $89$-axis, the states arrange
themselves in representations of the enhanced symmetry. There will be
additional restrictions from the $R_i$, but the discussion of
these restrictions is more efficiently left for later sections.
We turn now to the worldsheet consistency
conditions and the calculation of tadpoles.

\subs{Worldsheet Consistency and Tadpole cancellation}

We again use the notation of Ref. \cite{gimpol} where applicable. 
We denote the volumes of the three 2-tori as $V_i$,
and write $v_i=V_i/4\pi^2\alpha'$.
The volume of spacetime is denoted $V_4$, $v_4=V_4/(4\pi^2\alpha')^2$.
The cancellation of tadpoles for unphysical states is a
collaboration amongst the Klein bottle, the M\" obius strip
and the cylinder. We discuss each topology in turn.

\subsubsection{The Klein Bottle}

We compute the closed string trace 
\begin{equation}
\label{eq:kb_trace}
KB:\;\;\; \Tr\; {\Omega\over2} \left( {1+R_1\over 2}\right)
\left( {1+R_2\over 2}\right) 
\left( {1+(-1)^F\over 2}\right) q^{L_o+\tilde L_o}
\end{equation} 
Note that the orbifold part expands out to 
${1\over4}(1+R_1+R_2+R_3)$. We have $L_o=N_L+{1\over 2}\alpha' p_L^2-1/2$ with
$N_L=\sum r\alpha_{-r}\cdot\alpha_r+\sum r\psi_{-r}\cdot\psi_r$, etc.
and $p_{L,R}^2=\left( m_i/r_i\pm n_ir_i/\alpha'\right)^2$. $\Omega$ acts
trivially on $m$ and flips the sign of $n$. A projection $R$ changes
the sign of both (the appropriate) $m$ and $n$. The momentum integration
gives $(4\pi^2\alpha't)^{-2}$. $\Omega$ correlates left movers with
right movers and we have 
\begin{eqnarray}
\label{eq:closedomega}
\Omega\vac_{NS-NS}=-\vac_{NS-NS}\nonumber\\
(-1)^F\vac_{NS-NS}=-\vac_{NS-NS}\\
\Omega\vac_{R-R}=-\vac_{R-R}\nonumber\\
(-1)^F\vac_{R-R}=\pm\vac_{R-R}\nonumber 
\end{eqnarray} 
Since $\Omega\psi\Omega^{-1}=\tilde\psi$, 
$\Omega$ acts as $-(-1)^{F_L}$ on
all closed string states. So in the untwisted sector we get
contributions (``$\eta_3=-$'' in the language of Ref. \cite{polcai}): 
\begin{equation}
\begin{array}{rl} 
\Omega (-1)^F:& \left[{f_3\over f_1}(q)\right]^8
\prod_{j=1}^{3} M_j\\ 
\Omega R_i(-1)^F:& \left[{f_3\over f_1}(q)\right]^8 M_i
\prod_{j\neq i}W_j\nonumber 
\end{array}
\end{equation} 
from the NS-NS sector. (Here $q=e^{-2\pi t}$). This is to be
multiplied by ${1\over16}v_4\int {dt\over 2t^3}$. 
We have defined the momentum and winding factors for the $j^{th}$
torus
\begin{eqnarray} 
M_j=\left(\sum_n e^{-\pi tn^2/v_j}\right)^2\\ 
W_j=\left(\sum_m e^{-\pi tm^2v_j}\right)^2.\nonumber 
\end{eqnarray} 
The R-R sector gives the same
result with $f_3^8$ replaced by $-f_2^8$, with the contributions coming
from $\Omega$ and $\Omega R_i$ (those with $(-1)^F$ cancel, because of
the action on the R-R vacua). After appropriate rescalings and
resummations, we arrive at: 
\begin{equation} 
{v_4}\int_0^\infty {dt\over 2t^2} 
\left[ f_2(e^{-\pi/2t})\over f_1(e^{-\pi/2t})\right]^8 
\left[ \prod_i v_i\tilde M_i +
\sum_j v_j \tilde M_j\prod_{i\neq j} {\tilde W_i\over v_i} \right]
\end{equation} 
where 
\begin{eqnarray} 
\tilde M_i = \left(\sum_s e^{-\pi v_j s^2/t}\right)^2\\ 
\tilde W_i = \left(\sum_s e^{-\pi s^2/tv_i}\right)^2.\nonumber 
\end{eqnarray} 
Asymptotically
($t\rightarrow 0$), we get ($t=1/4\ell$): 
\begin{equation} 
32v_4\int {d\ell}\; 
\left( v_1v_2v_3+{v_1\over v_2v_3} +{v_2\over v_3v_1}+
{v_3\over v_1v_2}\right) .
\end{equation} 
Thus, the Klein bottle gives a tadpole for a 10-form
potential, as well as for 3 different 6-forms, each proportional to the
appropriate T-dual volume element. The three different
6-forms appearing here will ultimately be responsible for
the inclusion of three different Dirichlet 5-branes.

\subsubsection{The M\" obius Strip}

Here we evaluate \begin{equation}\label{eq:ms_trace} MS:\;\;\;
\Tr_{NS-R}\; {\Omega\over2} \left( {1+R_1\over 2}\right)\left(
{1+R_2\over 2}\right) \left( {1+(-1)^F\over 2} \right) q^{L_o}
\end{equation} over 99- and $5_i5_i$- states (the $\eta_3=-$ condition
of Ref. \cite{polcai} corresponds to NS states here). 
We have 
\begin{equation} 
L_o=\alpha' \vec p^2+\left[\sum r\alpha_{-r}\cdot\alpha_r+ \sum
r\psi_{-r}\cdot\psi_r-1/2\right]+ \left\{ 
\begin{array}{ll} 
m_i^2 \alpha' /r_i^2 & 99\\ 
n_i^2 r_i^2/\alpha' & 55 
\end{array}\right.
\end{equation} 
and the non-compact momentum integration now gives $(8\pi^2 t\alpha')^{-2}$.

$\Omega$ acts on oscillators as: 
\begin{eqnarray} 
\alpha_r&\rightarrow&
\pm e^{i\pi r}\alpha_r\left\{ \begin{array}{l}NN\\ DD\end{array}\right.
\\ 
\psi_r&\rightarrow& \pm e^{i\pi r}\psi_r\nonumber 
\end{eqnarray}
and as $e^{i\pi/2}$ on the open string vacuum 
(plus the action on Chan-Paton).

Consider first the 99-sector. 
%We have from the NS fermions:
%\begin{eqnarray} 
%\Omega:&\prod_r\left(1+e^{i\pi r}q^r\right)^8\nonumber\\ 
%\Omega (-1)^F:&-\prod_r\left(1-e^{i\pi r}q^r\right)^8\\ 
%\Omega R_i:&\prod_r\left(1+e^{i\pi r}q^r\right)^4
%\prod_r\left(1-e^{i\pi r}q^r\right)^4\nonumber\\ 
%\Omega R_i (-1)^F:&-\prod_r\left(1-e^{i\pi r}q^r\right)^4 
%\prod_r\left(1+e^{i\pi r}q^r\right)^4\nonumber 
%\end{eqnarray} 
%the products being over half-integers. The leading minus signs are 
%due to the $(-1)^F$ charge of the vacuum. 
%We see that the terms involving $R_i$ cancel in the
%99-sector. 
The oscillator sums involving $R_i$ turn out to cancel; 
the remaining terms can be written in the form
$iq^{1/2}\left[f_2f_4/f_1f_3(q)\right]^8$. This result
is to be multiplied by
${v_4\over 16}\int {dt\over (2t)^3} q^{-1/2}$, times the Chan-Paton and
momentum state sums, which are 
$i\Tr\gamma_{\Omega,9}^T\gamma_{\Omega,9}^{-1} \prod_i M'_i$. 
We define
\begin{eqnarray} 
M'_j=\left(\sum_n e^{-2\pi tn^2/v_j}\right)^2\\
W'_j=\left(\sum_m e^{-2\pi tm^2v_j}\right)^2.\nonumber 
\end{eqnarray}

In the $5_i5_i$-sector we find that the NS fermions give
\begin{eqnarray} 
\Omega:&\prod_r\left(1+e^{i\pi r}q^r\right)^4
\prod_r\left(1-e^{i\pi r}q^r\right)^4\nonumber\\ 
\Omega (-1)^F:&-\prod_r\left(1-e^{i\pi r}q^r\right)^4 
\prod_r\left(1+e^{i\pi r}q^r\right)^4\nonumber\\ 
\Omega R_i:&\prod_r\left(1+e^{i\pi r}q^r\right)^8\\ 
\Omega R_i (-1)^F:&-\prod_r\left(1-e^{i\pi r}q^r\right)^8\nonumber\\ 
\Omega R_j:&\prod_r\left(1+e^{i\pi r}q^r\right)^4 
\prod_r\left(1-e^{i\pi r}q^r\right)^4\nonumber\\ 
\Omega R_j (-1)^F:&-\prod_r\left(1-e^{i\pi r}q^r\right)^4
\prod_r\left(1+e^{i\pi r}q^r\right)^4\nonumber 
\end{eqnarray} 
for $r\in \ZZ+1/2$. The two
different terms above are from Neumann and Dirichlet boundary conditions
respectively. All but the $R_i$ terms cancel, and those are the same as
the 99- contribution; the boson contribution here is also the same as
the $\Omega$ terms from the $99$-sector.

The total result for the M\" obius strip then is 
\begin{equation} 
-{v_4\over 16}\int {dt\over (2t)^3} 
\left[ {f_2(q)f_4(q)\over f_1(q)f_3(q)}\right]^8 
\left\{ \Tr \gamma_{\Omega,9}^T\gamma_{\Omega,9}^{-1} \prod_i M'_i +
\sum_i \Tr\gamma_{\Omega R_i,5_i}^T\gamma_{\Omega R_i,5_i}^{-1} M'_i 
\prod_{j\neq i} W'_j\right\} 
\end{equation} 
Defining 
\begin{eqnarray} 
\tilde M'_i = \left(\sum_s e^{-\pi v_j s^2/2t}\right)^2\\ 
\tilde W'_i = \left(\sum_s e^{-\pi s^2/2tv_i}\right)^2,\nonumber 
\end{eqnarray} 
we obtain, by rescaling and resummation 
\begin{equation} 
\begin{array}{l} 
-{v_4\over 64}\int {dt\over t^2} 
\left[ {f_2f_4\over f_1f_3}\left( e^{-\pi/2t}\right)\right]^8\\ 
\;\;\;\;\;\;\;\;\times
\left\{ \Tr\gamma_{\Omega,9}^T\gamma_{\Omega,9}^{-1} 
\prod_i v_i\tilde M'_i +
\sum_i \Tr \gamma_{\Omega R_i,5_i}^T\gamma_{\Omega R_i,5_i}^{-1} 
v_i\tilde M'_i\prod_{j\neq i} {\tilde W'_j\over v_j}\right\}\nonumber
\end{array}
\end{equation} 
Finally, as $t\rightarrow 0$, we find ($t=1/8\ell$) 
\begin{equation} 
-{2v_4}\int {d\ell} 
\left[ v_1v_2v_3\;
\Tr \gamma_{\Omega,9}^T\gamma_{\Omega,9}^{-1} +
\sum_i v_i\prod_{j\neq i} {1\over v_j}\; 
\Tr \gamma_{\Omega R_i,5_i}^T\gamma_{\Omega R_i,5_i}^{-1}
\right] 
\end{equation}

\subsubsection{The Annulus}

Here we evaluate 
\begin{equation}
\label{eq:cyl_trace} 
Cyl:\;\;\;
\Tr_{NS-R}\; {1\over2} \left( {1+R_1\over 2}\right)
\left( {1+R_2\over 2}\right) \left( {1+(-1)^F\over 2} \right) 
q^{L_o} 
\end{equation} 
over 99-, $5_i5_j$-, $5_i9$- and $95_i$- states ($\forall i,j$). 
($\eta_3=-$ here corresponds to dropping the $(-1)^F$ terms). 
The Ramond sector carries an overall minus sign, which is the 
usual field theory sign for fermion loops.

The 99-sector: the fermions give: 
\begin{eqnarray} 
1 (NS):& \prod_r\left( 1+q^{r-1/2}\right)^8\\ 
1 (R):& -16\prod_r\left( 1+q^r\right)^8\nonumber 
\end{eqnarray} 
where $r\in\ZZ$. Including the bosons, we then find that
the oscillators contribute $q^{1/2}\left[ f_4(\sqrt{q})/
f_1(\sqrt{q})\right]^8$. The Chan-Paton factors are
$\Tr\gamma_{1,9}\cdot \Tr\gamma_{1,9}^{-1}$ and the momentum modes give
$\prod_i M'_i$. So we have so far 
\begin{equation}
\label{eq:ann_nn_a}
\begin{array}{rl} 
99:&\;\;\; {v_4\over 16}\int {dt\over (2t)^3} \left[
{f_4\over f_1}(\sqrt{q})\right]^8 
\left\{ \Tr\gamma_{1,9}\cdot\Tr\gamma_{1,9}^{-1}
\prod_i M'_i \right\} \\ 
&\rightarrow {1\over 32}v_4\int d\ell\; n_9^2\; v_1v_2v_3 
\end{array} 
\end{equation} 
with $n_9$ the number of 9-branes. The
$R_i$-operators in the trace give: 
\begin{eqnarray} 
R_i (NS):&\prod_r\left( 1+q^{r-1/2}\right)^4
\left( 1-q^{r-1/2}\right)^4\\ 
R_i (R):& 0\nonumber 
\end{eqnarray} 
and the bosons give $\prod_r (1-q^r)^{-4}(1+q^r)^{-4}$. 
The result is $4[f_3f_4/f_1f_2 (\sqrt{q})]^4$. 
Chan-Paton give $\Tr\gamma_{R_i,9}\cdot \Tr\gamma_{R_i,9}^{-1}$ and
there will be a factor of $M'_i$ from windings. The result is
\begin{equation}
\label{eq:ann_nn_b} 
\begin{array}{rl} 
99:&\;\;\;
{v_4\over 16}\int {dt\over (2t)^3} 
\left[ {f_3 f_4\over f_1f_2}(\sqrt{q})\right]^4 
\left\{ 4\sum_i M'_i\; \Tr\gamma_{R_i,9}\cdot
\Tr\gamma_{R_i,9}^{-1} \right\}\\ 
&\rightarrow {1\over 8}v_4 \int d\ell\; \sum_i v_i 
\Tr \gamma_{R_i,9}\cdot \Tr\gamma_{R_i,9}^{-1}.
\end{array} 
\end{equation}

The $5_i5_i$-sector: we have 4 NN and 4 DD bosons. We find
contributions: 
\begin{equation} 
\begin{array}{llc} 
1: & NS:\; & \prod_r \left( 1+q^{r-1/2}\right)^8\nonumber\\ 
& R: & -16 \prod_r \left( 1+q^{r}\right)^8\\ 
& Bos: & \prod_r \left( 1-q^{r}\right)^{-8}\nonumber\\ 
R_j: & NS:\; & \prod_r \left( 1+q^{r-1/2}\right)^4 
\left( 1-q^{r-1/2}\right)^4  \nonumber\\ 
&  R:   & 0\\ 
& Bos: & \prod_r\left( 1-q^{r}\right)^{-4} 
\left( 1+q^{r}\right)^{-4}\nonumber 
\end{array} 
\end{equation} 
Thus, the oscillators contribute for the unit operator 
$[f_4/f_1 (\sqrt{q})]^8$ and for the $R_i$-insertions
$+4[f_3f_4/f_1f_2 (\sqrt{q})]^4$. 
The unit operator piece gets multiplied by 
$$\sum_i M'_i\;\sum_{a,b\in 5_i}
(\gamma_{1,5_i})_{aa} (\gamma^{-1}_{1,5_i})_{bb}
\prod_{m_{j\neq i}}\sum_w
exp(-t(2\pi wr_j+X_a^{m_j}-X_b^{m_j})^2/2\pi\alpha').
$$ 
The $R_i$ operator pieces get just 
$\sum_i M'_i \sum_{I=1}^{16}
\Tr\gamma_{R_i,5_i}\cdot \Tr\gamma_{R_i,5_i}^{-1}$.

In the $5_i5_j$-sector ($i\neq j$) one has 4 ND bosons, 2 NN and 2 DD.
For the unit operator, NS gives $4\prod_r (1+q^{r-1/2})^4 (1+q^r)^4$ 
but
the Ramond sector is equal and opposite. For $R_\ell$ insertions, the 
NS
sector gives zero (because of cancellation between vacua). The Ramond
sector contributes only for $\epsilon_{ijk}R_k$. In this case, bosons
give $\prod_r (1+q^{r-1/2})^{-4} (1-q^r)^{-4}$ and the Ramond fermions
give $-4\prod_r (1-q^{r-1/2})^4 (1+q^r)^4$. The product is
$-[f_2f_4/f_1f_3 (\sqrt{q})]^4$.

The $5_i9$-sector is almost identical to the above case. The
only sector which contributes is the Ramond $R_i$ term, and the
oscillator parts again give $-[f_2f_4/f_1f_3 (\sqrt{q})]^4$.

\subsubsection{The Full Tadpole}

For brevity, let us drop terms proportional to $\Tr\gamma_R$, as these
are all zero as we will see in the next section. The tadpole is then
given by: 
\begin{equation} 
\begin{array}{l} 
{1\over 32}v_4\int {d\ell}\;
	\left\{ 
		v_1v_2v_3\left( 
		32^2-64 \Tr\gamma_{\Omega,9}^T\gamma_{\Omega,9}^{-1} +n_9^2 
	\right)\right. \\ 
+	\left.\sum_i v_i\prod_{j\neq i} {1\over v_j}\; 
		\left( 
		32^2 -64\Tr
		\gamma_{\Omega R_i,5_i}^T\gamma_{\Omega R_i,5_i}^{-1}
		+n_{5_i}^2\right)
% 
%+\sum_i v_i\left( 4\Tr\gamma_{R_i,9}\cdot\Tr\gamma_{R_i,9}^{-1}
%+\sum_{I_i=1}^{16} \Tr\gamma_{R_i,I_i}\cdot\Tr\gamma_{R_i,I_i}^{-1}
%\right) 
% etc. 
% 
\right\} 
\end{array} 
\end{equation} 
Tadpole cancellation then implies 
\eqn{tdplcycla}
{	\gomt{,9}=+\gom{,9},}
\eqn{tdplcyclb}
{	\gomrt{i}{,5_i}=+\gomr{i}{,5_i}.} 
and a total of 32 9-branes plus 32 of each of three types of $5$-brane.

\subs{Chan-Paton Representation of $\Omega,R_i$} 
\label{sec:reprCP}

The basic orientifold group is defined, in part, by the following  
generators and relations:
\eqn{ora}
{(\Omega R_i)^2=1,\ \Omega^2=1,}
\eqn{orab}
{(\Omega R_i)(\Omega R_j)(\Omega R_k)\Omega=1,\ \;\; 
(\Omega R_i)\Omega(\Omega R_i)\Omega=1. }
for $j\neq k\neq i$.
We would like the $\gamma$-matrices, in
conjunction with the action of the orientifold group on the bulk CFT, to
form a projective representation of the orientifold 
group\fnote{*}{Although one can perhaps relax this condition.}. 
The bulk CFT contribution is easy to calculate, as explained 
in \cite{gimpol}, leaving us with what 
might be considered strange phase differences between different sectors. 

The relations (\ref{eq:ora}),(\ref{eq:orab}) imply the
following relations
\eqn{oraa}
{\gamma_{\Omega R_i}{\gamma_{\Omega R_i}^{-1}}^T=c_i(s),\
\gamma_{\Omega}{\gamma_{\Omega}^{-1}}^T=c(s)}
\eqn{oraab}
{\gamma_{\Omega R_i}{\gamma_{\Omega R_j}^{-1}}^T\gamma_{\Omega R_k}
{\gamma_{\Omega}^{-1}}^T=\rho_{ijk}(s),
\gamma_{\Omega R_i}{\gamma_{\Omega}^{-1}}^T\gamma_{\Omega R_i}
{\gamma_{\Omega}^{-1}}^T=\rho_i(s)}
where the factors on the RHS differ from sector to sector (the sectors
 are labeled by $s$).

We will analyze in some detail a few of these phases, and just list 
the others. We will argue that in addition to 
(\ref{eq:tdplcycla}), (\ref{eq:tdplcyclb})
we must have co-cycles of the following form
\eqn{mrcycla}
{	\gomt{,5_i}=-\gom{,5_i},}
\eqn{mrcyclb}
{	\gomrt{i}{,9}=-\gomr{i}{,9},\
	\gomrt{i}{,5_j}=-\gomr{i}{,5_j}.
} 
To show this, we need to examine all the
mixed sectors, and in fact the problem is over-determined.

Let us begin with $\gamma_\Omega$. The method is explained in
Refs. \cite{gimpol},\cite{mdgm}. 
Consider first the product of two vertex operators 
in the $95_i$ sector. We obtain, in either the $99$- or the
$5_i5_i$-sector, the product of two complex fermions, both of which 
are either NN or DD. Each of these rotates by a phase $\pm i$ under
$\Omega$, and so the two contribute $-1$ to the action of $\Omega$ in
the $99$ or $5_i5_i$ sectors. This corresponds to $\Omega^2=-1$
on the oscillators of the $95_i$ sector and hence we
obtain\cite{gimpol} 
the $-1$ in eq. (\ref{eq:mrcycla}). Note that this 
is consistent also with the $5_i5_j$-sector: taking the product of two
such vertex operators we obtain, in either the $5_i5_i$- or 
$5_j5_j$-sector, a product of two complex fermions, one DD and the 
other NN. This leads to the same phase for all $i$, as in 
(\ref{eq:mrcycla}).

To show the consistency of (\ref{eq:mrcyclb}), we need to look at 
$R_i\Omega$ in each of four sectors. Again we consider the product
of two vertex operators of a given sector.
\begin{description} 
\item [$\bullet$ $95_i$:] Taking the product of two such operators, we
obtain, in the $99$- or $5_i5_i$-sector, the product of two fermions as
discussed above. Both change sign under $R_i$, giving a net $-1$ sign 
between eq. (\ref{eq:tdplcyclb}) and the first of eqs.
(\ref{eq:mrcyclb}).

\item [$\bullet$ $5_i5_j$:] Here, only one of the two fermions will
change sign under $R_i$, giving a net $-1$ sign between eq.
(\ref{eq:tdplcyclb}) and the second of eqs. (\ref{eq:mrcyclb}).

\item [$\bullet$ $95_j$ ($i\not=j$):] In the 99 ($5_j5_j$) we obtain the
product of two complex NN(DD) fermions; one of
them is flipped by $R_i$ giving a relative phase of $+1$ between 
the 9- and $5_j$-sectors, consistent with eq. (\ref{eq:mrcyclb}).

\item [$\bullet$ $5_j5_k$ ($j,k\not=i$):] In the $5_j5_j$ ($5_k5_k$) we
obtain the product of an NN and a DD fermion; both are flipped by $R_i$
giving a relative phase of $+1$ between 
the $5_j$- and $5_k$-sectors, consistent with eq. (\ref{eq:mrcyclb}). 
\end{description} 

Thus the conditions (\ref{eq:mrcycla}),(\ref{eq:mrcyclb}) are 
implied by (\ref{eq:tdplcycla}),(\ref{eq:tdplcyclb}) and are, fortunately,
self consistent.

We still need to check the other relations that define the 
orientifold group, and determine the relative phase between different
sectors. The phase differences can be read from the explicit  
solution in the next section, which is unique up to unitary transformations
on the Chan-Paton factors. At this stage we will suffice in presenting
all the relevant information in the following table.
\begin{center}
\begin{tabular}{r|ccccc|} 
& $\Omega^2$ & $R_1^2$ & $R_2^2$ & $R_3^2$ & $R_1R_2R_3$ \\ \hline 
99   & $+1$ &  $+1$ & $+1$ & $+1$ & $+1$ \\
$5_i5_i$ & $+1$ &  $+1$ & $+1$ & $+1$ & $+1$ \\ 
$95_1$ & $-1$ & $+1$ & $-1$ & $-1$ & $-1$ \\ 
$95_2$ & $-1$ & $-1$ & $+1$ & $-1$ & $-1$ \\
$95_3$ & $-1$ & $-1$ & $-1$ & $+1$ & $+1$ \\ 
$5_15_2$ & $+1$ & $-1$ & $-1$ & $+1$ & $+1$ \\ 
$5_35_1$ & $+1$ & $-1$ & $+1$ & $-1$ & $-1$ \\
$5_25_3$ & $+1$ & $+1$ & $-1$ & $-1$ & $-1$ \\ \hline 
\end{tabular}
\end{center}
We have again looked at mixed sectors. As before this determines the
difference in the phases between the different unmixed sectors, and as
before the problem is over-determined. To obtain the phase difference
between, say, some $5_i5_i$ sector and, say, $99$ sector, pick the 
appropriate line in the table and multiply the appropriate phases in the row.

\subsubsection{A definite choice} 
\def\eps{\varepsilon}

Let us now make a definite choice for the $\gamma$'s. Define
\begin{equation} 
M_i= \left\{
\pmatrix{ 0&I \cr -I&0}, 
\pmatrix{ -\eps &0\cr 0&\eps},
\pmatrix{ 0& \eps\cr \eps &0 }\right\},
\end{equation} 
\begin{equation} 
D=\pmatrix{ 0&-\eps\cr \eps&0 } 
\end{equation} 
where $\eps=i\sigma_2$ and each block is understood to be
a direct product with the $k$-dimensional identity matrix,
$k\leq 32$ being the number of concident $p$-branes of the
appropriate type.  These matrices satisfy $M_i^2=-I$,
$M_iM_j=\epsilon_{ijk}M_k$ and $M_i^T=M_i^{-1}$. 
It is useful to also define 
$$N_i\equiv DM_i=\left\{\pmatrix{\eps&0\cr 0&\eps},
\pmatrix{0&I\cr I&0}, \pmatrix{I&0\cr0&-I} \right\}.$$

\newcommand{\cocy}[1]{{\cal C}_{#1}}

We will work in an enlarged matrix notation consisting of blocks
corresponding to $9, 5_1, 5_2, 5_3$-branes. It is convenient to define
cocycles $\cocy{1}={\rm diag}( I,-I,I,I )$, etc., and
$\cocy{0}=-\cocy{1}\cocy{2}\cocy{3}={\rm diag}( -I,I,I,I )$.

Collecting the results of the previous section, we have: 
\begin{eqnarray} 
\gomt{}&=&-\cocy{0} \gom{}\\
\gomrt{i}{}&=&-\cocy{i} \gomr{i}{} 
\end{eqnarray} 
and
\begin{eqnarray}
\label{eq:cocyclecons}
\gomr{i}{}\gomr{j}{}^{-T}\gomr{k}{}\gom{}^{-T} & = &
\cocy{0}\cocy{3}\epsilon_{ijk} I \;\;\; ({\rm for}\; i\neq j\neq k)\\
\gomr{i}{}\gom{}^{-T}\gomr{i}{}\gom{}^{-T} & = & \cocy{0}\cocy{i} I\\
\gamma_{R_i}\gamma_{R_j}\gamma_{R_k} & = &
\cocy{0}\cocy{3}\epsilon_{ijk} I\\ 
\gamma_{R_i}\gamma_{R_i} & = & \cocy{0}\cocy{i} I 
\end{eqnarray}

There are many choices consistent with the above constraints. A
particular example, consisting entirely of orthogonal matrices is
\begin{center} 
\begin{tabular}{l|rrrrrrr|}
&$\gom{}$&$\gomr{1}{}$&$\gomr{2}{}$&$ \gomr{3}{}$&$\gamma_{R_1}
$&$\gamma_{R_2}$&$\gamma_{R_3}$\\ \hline 
$9$ & $I$ & $M_1$ & $M_2$ &
$M_3$ & $-M_1$ & $-M_2$ & $-M_3$ \\ 
$5_1$ & $N_1$ & $-D$ & $M_2$ &
$M_3$ & $-M_1$ & $N_3$ & $-N_2$ \\ 
$5_2$ & $N_1$ & $M_3$ & $-D$ &
$M_2$ & $-N_2$ & $-M_1$ & $N_3$ \\ 
$5_3$ & $N_1$ & $M_2$ & $M_3$ &
$D$ & $N_3$ & $-N_2$ & $M_1$\\ \hline 
\end{tabular} 
\end{center} 
This choice satisfies $\gamma_{R_i}=\cocy{0}\gomr{i}{}\gom{}$.
and
\begin{equation}
\gamma_{R_i}^T=\cocy{0}\cocy{i}\gamma_{R_i}
\end{equation}
In fact the choice of matrices given in the table is essentially
unique: one can show that the available gauge freedoms allow
one to put any consistent choice in this form. This is not a trivial
statement; a-priori we can choose the factors $\rho_{ijk},\ \rho_i$ in
the 99 sector to be whatever we want (and then they are detemined in the
other sectors) but this does not seem to be the case. If we start with
any values for these constants other then the ones that correspond to
this solution we cannot solve all the constraints. 

\subs{The Spectrum}

In this section we examine the spectrum and interactions dictated by the
algebraic and tadpole cancellation conditions found above. The conditions
on Chan-Paton matrices of individual states 
were outlined in Section 2.2. We consider each
sector in turn.

\subsubsection{99-sector}

We have $\gom{,9}=I$, $\gomr{i}{,9}=\gamma_{R_i,9}=-M_i$. Thus from eqs.
(\ref{eq:open_nn}), the 99-gauge boson Chan-Paton factor 
$\lambda^{(0)}$ is antisymmetric and satisfies 
$\lambda^{(0)}=-M_i \lambda^{(0)} M_i$. The
solution to these conditions is ($A$=antisym., $S$=sym.) :
$$\lambda^{(0)}=\pmatrix{  A  & S_1& S_2& S_3\cr -S_1& A  &S_3& -S_2\cr
-S_2& -S_3& A  &S_1\cr -S_3&S_2& -S_1& A }$$ which is the adjoint
representation of $Sp(8)$. This may be broken to smaller groups by the
addition of Wilson lines.

The matter fields $\lambda^{(i)}$  are 
\begin{equation}
\lambda^{(1)}=\pmatrix{  A_1& A_2 & S  & A_3\cr 
						 A_2&-A_1 &-A_3& S  \cr
						  -S&-A_3 & A_1& A_2\cr  
						 A_3& S   & A_2&-A_1 };\;\;\;
\lambda^{(2)}=\pmatrix{  A_3& S   & A_1& A_2\cr 
 						-S  & A_3 & A_2&-A_1\cr 
 						 A_1& A_2 &-A_3&-S  \cr 
 						 A_2&-A_1 & S  &-A_3 }; 
\end{equation} 
\begin{equation}
\lambda^{(3)}=\pmatrix{  A_1& A_2& A_3& S  \cr 
						 A_2&-A_1&-S  & A_3\cr
						 A_3& S  &-A_1&-A_2\cr 
						 -S & A_3&-A_2& A_1 } 
\end{equation}
\newcommand{\tensa}{\tableaux{\yboxs\cry\yboxs\cry}} 
These are six bosons (giving 3 chiral multiplets) in the 
$\tensa={\bf 120}$ of $Sp(8)$.

\subsubsection{$5_i5_i$-sector}
\newcommand{\tlam}[1]{\tilde\lambda^{(#1)}}
\newcommand{\tenss}{\tableaux{\yboxs\yboxs\cry}}

First take the $5_i$ which are at fixed points of $R_i$. For example, for
$i=1$, we have, from Section 2.2, the conditions 
$$
\lambda^{(0)T}=N_1\lambda^{(0)}N_1;\;\;\;
\lambda^{(0)}=-M_{1}\lambda^{(0)}M_{1};\;\;\;
\lambda^{(0)}=+N_{2,3}\lambda^{(0)}N_{2,3} 
$$ 
It is convenient to define $\tlam{A}=N_1\lambda^{(A)}$;
if $4k$ $5_i$-branes are at the same fixed point of $R_i$, we
obtain an $Sp(k)$ gauge group, with 
$$
\tlam{0}=\pmatrix{  S& 0\cr 0& S}.
$$ 
where the blocks are $2k$-dimensional ($k\leq 8$).

Applying the conditions stated in Section 2.2, we find that $\tlam{1}$
is symmetric, and $\tlam{2,3}$ antisymmetric. Solutions are:
\begin{equation}
\tlam{i}=\left\{
\pmatrix{0& A\cr -A&0} , \pmatrix{A&0\cr 0&-A}, \pmatrix{0& A\cr A&0}
\right\}
\end{equation} 
Thus we again get 3 chiral multiplets in the
$\tableaux{\yboxs\cry\yboxs\cry}$. For all $5_i$-branes at the same
fixed point of $R_i$, we see that the spectrum is completely T-dual.

Now consider moving 5-branes away from the fixed points. First suppose
we move the branes away from the fixed point of $R_i$ along the fixed
direction of $R_j$ (refer again to Fig. 1). For definiteness, consider a
$5_1$-brane on the $6,7$-axis (i.e. along the fixed line of $R_2$).
$R_1$ and $R_3$ map a brane to its image and so place no restrictions.
Thus we have the conditions: 
\begin{eqnarray}
\lambda^{(0,1)T}=+N_1\lambda^{(0,1)}N_1;\;\;
\lambda^{(2,3)T}=-N_1\lambda^{(2,3)}N_1\\
\lambda^{(0,2)}=+N_3\lambda^{(0,2)}N_3;\;\;
\lambda^{(1,3)}=-N_3\lambda^{(1,3)}N_3 
\end{eqnarray} 
With $\tlam{A}$ defined as above, 
$\tlam{0,1}$ are symmetric and $\tlam{2,3}$ antisymmetric. 
If, as above, we started with $4k$ 5-branes at the fixed point,
we end up with at most $2k$ 5-branes along the $67$-axis (their
images accounting for the other $2k$). We then find 
\begin{equation}
\label{eq:outlamo} 
\tlam{0}=     \pmatrix{ S_1& S_2\cr S_2& S_1};\;\;\; 
\tlam{1}=     \pmatrix{  S& A\cr -A&-S}; 
\end{equation}
\begin{equation}
\label{eq:outlamt} 
\tlam{2}=     \pmatrix{ A_1& A_2\cr A_2& A_1};\;\;\; 
\tlam{3}=     \pmatrix{  A& S\cr -S&-A};   \nonumber
\end{equation} 
where the blocks of these matrices 
are at most $k\times k$-dimensional.
$\tlam{0}$ is the adjoint of $Sp(k/2)\times Sp(k/2)$. Note the 
difference in
form between $\tlam{2}$ and $\tlam{1,3}$; this is because the 
motion of
$5_1$-branes along the $R_2$-fixed line is T-dual to turning on the
$\lambda_2$ of the 99-sector. We expect then that $\lambda^{(2)}$
is special, being identified with the Higgs mode.
Each of the $Sp(k/2)$ thus has  matter in
$\tenss+2\times\tensa$. Note that these fit appropriately into vector
and hypermultiplets of $N=2$ supersymmetry. As we will see later, a
detailed understanding of this spectrum in terms of Higgsing of the
previous case relies crucially on the presence of the tree-level
superpotential. Note also that this all occurs only for even $k$ (only
a group of eight 5-branes may move off of a fixed point); thus
the moduli space breaks up into disconnected sectors, in a fashion 
similar to that of Ref. \cite{gimpol}.

If we move the 5-branes away from fixed planes entirely, there are three
images, and the $R_i$ give no condition on states. In the $55$-sector,
$\Omega$ gives (for $5_1$ branes) $\lambda^{(0,1)}=
N_1\lambda^{(0,1)T}N_1$ and $\lambda^{(2,3)}= -N_1\lambda^{(2,3)T}N_1$.
Proceding as above, we may define $\lambda^{(A)}=N_1\tlam{A}$; the
$\tlam{0,1}$ will be symmetric and the $\tlam{2,3}$ antisymmetric, but
otherwise unconstrained. We thus find an $Sp(k/2)$ gauge group, with
matter in $\tenss+2\times\tensa$. In Higgsing from $Sp(k/2)\times Sp(k/2)$,
the $\tenss$ of one group is eaten, while the $\tensa$'s, as we discuss
later, are made massive through the superpotential. 
Note that this spectrum can be arranged into multiplets of $N=2$ 
supersymmetry. The operators which
appear in the (renormalizable) superpotential are also consistent
 with the enhanced supersymmetry. In fact we do not expect this
to be an exact symmetry; it merely reflects the fact that 
isolated $5$-branes act like those of Ref. \cite{gimpol}.

\subsubsection{$95$- and $55$-sectors}

Let us now check $95$ and $5_i5_j$ states. If we have $4k_9$ 9-branes
and $4k_{5_1}$ $5_1$-branes at a fixed point of $R_1$, we find
projections 
\begin{eqnarray} 
\lambda=-M_1\lambda M_1\\ 
\lambda=-\eta\; M_2\lambda N_3 = +\eta\; M_3\lambda N_2
\end{eqnarray} 
where $\eta=(2s_4)i$, as in eq. (\ref{eq:fivninproj}). 
This has solution: 
$$ \lambda=\pmatrix{ m_1& m_2& m_3& m_4\cr 
		-\eta m_1&-\eta m_2& \eta m_3&\eta m_4\cr 
			-m_3& -m_4& m_1& m_2\cr 
			-\eta m_3& -\eta m_4& -\eta m_1& -\eta m_2 } 
$$ 
where $m_i$ are
arbitrary $k_9\times k_{5_1}$ matrices. This gives one chiral 
superfield in the 
$({\bf 2k_9, 2k_{5_1}})= (
\tableaux{\ybox\cry},\tableaux{\ybox\cry})$ 
of $Sp(k_9)\times Sp(k_{5_1})$.

If the $5_1$-brane is moved out along the fixed line of $R_2$, we have
only the constraint from $R_2$: $\lambda=-\eta M_2\lambda N_3$. 
This will transform as
$(\tableaux{\ybox\cry},\tableaux{\ybox\cry},1) \oplus
(\tableaux{\ybox\cry},1,\tableaux{\ybox\cry})$ 
of the $Sp(k_9)\times Sp(k_{5_1}/2)\times Sp(k_{5_1}/2)$ 
gauge group. 

If we move the 5-branes away from fixed planes entirely, there are no
constraints, and again we find
$(\tableaux{\ybox\cry},\tableaux{\ybox\cry})$ of the unbroken gauge
group. Note that since the corresponding Higgsing does not give vevs to
these fields, it must be that the extra fields are made massive by the
superpotential. Indeed this will be found in the next section.

As discussed in Section 2, when two different $5$-branes 
intersect\cite{vbs} one another, there are massless $5_i5_j$
states. These will transform in the 
$(\tableaux{\ybox\cry},\tableaux{\ybox\cry})$ of the 
appropriate groups. Again, the dynamics of these states
depend crucially on the existence of a superpotential,
and so we turn to that now.

\subs{The Superpotential}

We have indicated in several places above the importance of
the tree-level superpotential in the proper understanding
of the symmetry breaking phenomena discussed above.
We study this in some detail in this section, concentrating on
couplings of open string modes only, which are relevant for
the discussion of Higgsing.

The matter multiplets of open string states organize themselves into
chiral multiplets $z_i=\phi^{2m_i}+i\phi^{2m_i-1}$, one per $T^2$. 
In addition, in the $59$ and $5_i5_j$ sectors, there are
states which we will label simply by $z$, with a superscript
to identify the appropriate sector. The
superpotential should be invariant under automorphisms of the
orientifold group.

We consider here only the renormalizable part of the superpotential 
connecting open string states. We will determine this by consideration 
of three-point disc amplitudes involving two Ramond states and 
one NS state. If such an amplitude is non-vanishing, it can of course 
be interpreted as a term in the superpotential. We can take 
vertex operators in the canonical ghost pictures: 
\begin{equation} 
\begin{array}{rl} 
V_{-1/2}^{(j)}
&=u^{(j)}_\alpha S^\alpha e^{-\phi/2} e^{i\vec w_j\cdot \vec H}
e^{ik\cdot X}\\ 
V_{-1}^{(j)} &=e^{-\phi}e^{i\vec v_j\cdot\vec H} e^{ik\cdot X} 
\end{array} 
\end{equation} 
In order for a term $W=z_iz_jz_k$ to
exist, we must have $\vec w_i +\vec w_j+ \vec v_k=0$. Traces over
Chan-Paton factors will be understood. We identify the $w_i$ via their
transformation properties under the $R_j$. 
\begin{description} 
\item [$\bullet$ $ 99$:] In the 99 sector, we have 
$\vec w_1=\left(\pm\half,\mp\half,\pm\half\right)$, 
$\vec w_2=\left(\mp\half,\pm\half,\pm\half\right)$, 
$\vec w_3=\left(\pm\half,\pm\half,\pm\half\right)$, 
$\vec v_1=\left(\pm 1,0,0\right)$, 
$\vec v_2=\left(0,\pm 1,0\right)$, and 
$\vec v_3=\left(0,0,\mp 1\right)$. 
We deduce that there is a term
$$W=z_1^{99}z_2^{99}z_3^{99}$$.

\item [$\bullet$ $ 5_19$:] The Ramond state has 
$\vec w_1=\left(\pm\half,0,0\right)$. 
There are thus additional terms of the
form $$W=z_i^{99}z^{95_i}z^{5_i9}$$.

\item [$\bullet$ $ 5_i5_i$:] The vectors will be similar to 
those of the 99-sector. Thus we expect
$$W=z_1^{5_i5_i}z_2^{5_i5_i}z_3^{5_i5_i}+z_i^{5_i5_i}z^{5_i9}z^{95_i}$$

\item [$\bullet$ $5_i5_j$:] When 5-branes cross, there will be
additional massless states, and there can be couplings of the form
$$W=\epsilon_{ijk} z^{5_i5_j}z^{5_j5_i}z_k^{5_i5_i}
+z^{5_15_2}z^{5_25_3}z^{5_35_1} +z^{5_i5_j}z^{5_j9}z^{95_i}.$$

\end{description}

\subs{Further Comments on Higgsing}

With an understanding of the superpotential, we can return to the
discussion of Higgsing related to moving 5-branes away from fixed
points. Consider again $32$ $5_1$-branes at a fixed point of $R_1$.
Moving some of these away is T-dual to turning on Wilson lines in the
9-brane sector, or equivalently, to giving a vev to one of the
antisymmetric tensor fields. Thus we expect that moving 5-branes should
have a low energy description in terms of Higgsing. Consider moving
$2k_5$ $5_1$-branes out along the fixed line of $R_2$. This corresponds
to turning on the antisymmetric tensor field $z_2^{5_15_1}$, and as we
have seen in previous sections, the gauge group is broken to
$Sp(8-k_5/2)\times Sp(k_5/2) \times Sp(k_5/2)$, the first factor arising from
those branes still at the $R_1$ fixed point. As far as $z_2^{5_15_1}$ is
concerned, the physics is simple: some of its components are eaten, and
the remaining components appear as in eq. (\ref{eq:outlamt}). Certain
components of the other $5_15_1$ modes are given mass through the
superpotential 
$$
W=\langle z_2^{5_15_1}\rangle z_1^{5_15_1}z_3^{5_15_1}
$$ 
leading to the remaining fields in eqs. (\ref{eq:outlamo}),
(\ref{eq:outlamt}). As well, if there are massless $5_15_j$ ($j\neq 1$)
modes, parts of these will be given mass by the superpotential as well.
This is completely sensible, and the situation is displayed
in Figure 4. The vev that we have been discussing corresponds
to moving the $5_1$-brane vertically, leaving the $5_3$-brane
in place, which clearly gives mass to those open string modes
stretching between the two. 
\efig{2in}{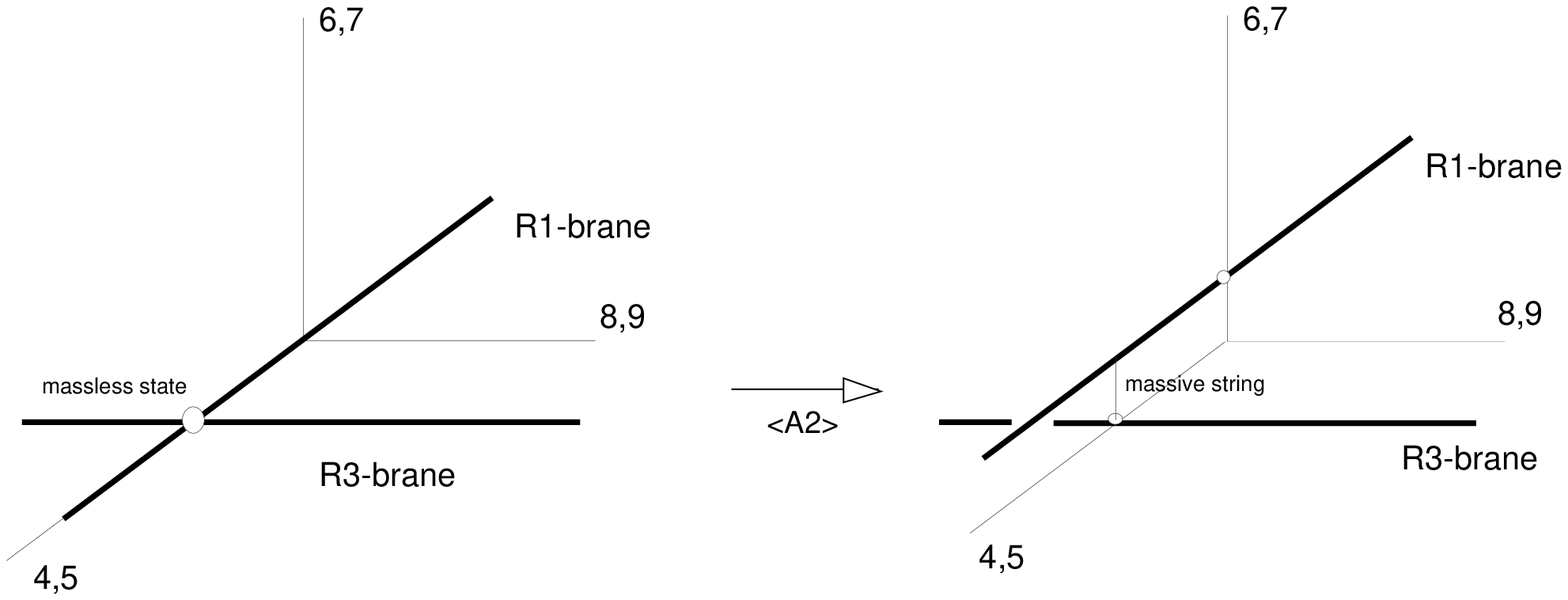}{Mass of $5_i5_j$ states
states reproduced by superpotential through Higgsing.} 

Note however that if the two different $5$-branes are 
moved together up the 67-axis in Figure 4, we
should see from the superpotential that the $5_15_3$-state is not made
massive. Indeed, this situation corresponds to turning on both $z_2^{5_15_1}$
and $z_2^{5_35_3}$. The relevant terms in the
superpotential are 
$$
W=\left(\langle z_2^{5_15_1}\rangle-\langle
z_2^{5_35_3}\rangle \right) z^{5_15_3}z^{5_35_1}
$$
When the two vevs are turned on equally, the $5_15_3$-states 
are not lifted. 

It is clear from these examples that the motion of Dirichlet
$5$-branes is completely equivalent to Higgsing provided
the superpotential is taken into account.

\subs{Conclusions}

In this paper, we have presented a four-dimensional model
with $N=1$ supersymmetry which is an orbifold of Type I
superstrings, and have concentrated on some of the technology that goes into 
such a construction.  A fully consistent picture emerges if three
types of Dirichlet 5-branes are included, all of which fill
the four space-time dimensions. We have confined ourselves to 
a tree-level analysis of (what are from the low energy point of view),
Higgsing phenomena. Clearly many
outstanding issues remain particularly those related to
non-perturbative issues and duality; we will return to these
in a future publication\cite{mbrltwo}. 

We would like to thank T. Banks and M. Douglas for helpful discussions.
Research supported in part by the U.S. Department of Energy, 
contract DE-FG05-90ER40599.

\pagebreak 

\end{document}